\documentclass[english,aps,preprintnumbers,amsmath,amssymb,nofootinbib,twocolumn]{revtex4-1}

\usepackage[latin1]{inputenc}
\usepackage{graphicx}
\usepackage{color}

\usepackage{color}

\usepackage{bm}
\usepackage{amsmath}

\usepackage{dsfont}

\newcommand{\be}{\begin{eqnarray}}
\newcommand{\ee}{\end{eqnarray}}

\newcommand{\nn}{\nonumber }

\newcommand{\Nf}{N_{\text{f}}}

\newcommand{\pat}{\partial_t}
\newcommand{\Eqref}[1]{Eq.~\eqref{#1}}

\newcommand{\fslash}{\hspace{-0.1ex} \slash }
\def\slash#1{\setbox0=\hbox{$#1$}               
   \dimen0=\wd0                                 
   \setbox1=\hbox{/} \dimen1=\wd1               
   \ifdim\dimen0>\dimen1                        
      \rlap{\hbox to \dimen0{\hfil/\hfil}}      
      #1                                        
   \else                            
      \rlap{\hbox to \dimen1{\hfil$#1$\hfil}}   
      /                                         
   \fi}                                         %

\newcommand{\psibar}{\bar{\psi}}
\newcommand{\spinor}{\psi_{j}}
\newcommand{\barspinor}{\bar{\psi}_{j}}
\newcommand{\altbarspinor}{\bar{\psi}_{i}}
\newcommand{\altspinor}{\psi_{i}}

\usepackage{babel}
\makeatother

\begin{document}

\title{Renormalization Group Study of Magnetic Catalysis in the $3d$ Gross-Neveu Model}
\author{Daniel D.~Scherer and Holger Gies}
\affiliation{Theoretisch-Physikalisches Institut, Friedrich-Schiller-Universit\"at Jena, 
Max-Wien-Platz 1, D-07743 Jena, Germany}

\begin{abstract}
Magnetic catalysis describes the enhancement of symmetry breaking quantum fluctuations 
in chirally symmetric quantum field theories by the coupling of fermionic degrees of 
freedom to a magnetic background configuration. We use the functional renormalization group 
to investigate this phenomenon for interacting Dirac fermions propagating in (2+1)-dimensional
space-time, described by the Gross-Neveu model. We identify pointlike operators up to quartic fermionic 
terms that can be generated in the renormalization group flow by the presence of an external magnetic field. 
We employ the beta function for the fermionic coupling to quantitatively analyze the field dependence of the induced spectral gap.
Within our pointlike truncation, the renormalization group flow provides a simple picture for magnetic catalysis.  
\end{abstract}

\maketitle

\section{Introduction}\label{sec:intro}

In the past 20 years it has become evident that in certain planar condensed matter model
systems defined on some lattice, the low-energy physics can efficiently be described in terms of Dirac fermions. These ``relativistic''
degrees of freedom typically arise by linearizing a given dispersion relation around a special set of
points in the Brillouin zone, where either a linear band crossing is realized or the single-particle
gap function vanishes at isolated points across a given Fermi surface.

The most prominent and well-investigated example is probably the Hubbard model on the honeycomb lattice~\cite{Herbut:2009qb,Gusynin:2007ix}. 
This model describing the hopping of fermions on a hexagonal lattice with possibly strong on-site interactions serves as a 
minimal model for the electronic correlations in monolayer graphene and similar systems. Once the naive continuum
limit has been taken, one can expect a faithful representation of the low-energy sector by a quantum field theory for Dirac fermions moving in $(2+1)$-dimensional space-time in the
proximity to a second order quantum phase transition. The lattice symmetries of the Hubbard model
get replaced by an ``emergent'' Lorentz symmetry and the free part of the Dirac action is invariant under continuous
chiral transformations. Note that for the chiral symmetry to be realized, a reducible four-component representation
of the Dirac algebra is necessary, which can be formed from the direct sum of two inequivalent two-component representations. In the case of an irreducible representation, 
the chiral symmetry effectively corresponds to part of the flavor symmetry.
In the following, we will consider the chirally invariant case. This asymptotic chiral symmetry is also realized for example in $d$-wave 
superconductors~\cite{Vojta:2000zz,Khveshchenko:2000} or quantum spin-liquid phases~\cite{Fradkin:1991nr,Wen:2004ym}. In both cases, however, 
a symmetry of the underlying parent state is already assumed to be broken, in
contrast to the case of the Hubbard model on the honeycomb lattice. Another
situation where no symmetry breaking is necessary are topological insulators and superconductors, which support Dirac fermions as surface states \cite{Hasan:2007,Hasan:2010xy}.

Once interactions are included in such a low-energy effective quantum field theory, we expect that 
the chiral symmetry is spontaneously broken as soon as the associated coupling exceeds a critical
value. The fluctuation-driven interactions compatible with the symmetries determine in which channel the formation of a condensate,
i.e., a finite expectation value of an order-parameter field, can take place. In this paper, the focus of
our investigation is on the Gross-Neveu model with a reducible representation of the Dirac algebra and a purely pointlike
scalar-scalar interaction. This theory realizes a system with discrete chiral symmetry. Quantum phase transitions falling 
into the Gross-Neveu universality class for two fermion flavors $N_{\mathrm{f}}=2$ are, for example the excitonic~\cite{Khveshchenko:2001zz} and antiferromagnetic instabilities~\cite{Herbut:2006cs}
in graphite and graphene, respectively. The universal properties of secondary $d$- to $d+\mathrm{i}s$-wave pairing transitions in nodal $d$-wave superconductors
are also expected to be described by the Gross-Neveu
model~\cite{Vojta:2000zz,Khveshchenko:2000}. Chiral symmetry breaking is known
to occur for all flavor numbers $N_{\mathrm{f}}$~\cite{Braun:2010tt}, 
whereas in models with continuous chiral symmetry and interactions in a vector-vector channel such as the Thirring model~\cite{Gies:2010st}, a chiral condensate is stable only up to a critical flavor number.  

Remarkably, when interacting planar Dirac fermions are subject to an external perpendicular magnetic field, symmetry breaking accompanied by
the generation of a mass gap for the fermionic degrees of freedom occurs for all values of the interaction strength~\cite{Gusynin:1994va}. Indeed, this effect known by
the name of magnetic catalysis is not particular to models defined in three space-time dimensions, but has also been investigated in four dimensional
quantum field theories~\cite{Gusynin:1994xp,Gusynin:1995nb}. However, what is special about three space-time dimensions is the fact that even in the absence of interactions a finite
value of the chiral condensate is generated by applying an external magnetic field~\cite{Gusynin:1994va}. So also for a free system the (continuous) chiral symmetry is spontaneously broken. This phenomenon
is sometimes referred to as a `quantum anomaly'~\cite{Gusynin:2005pk}, since it unexpectedly leads to the generation of a non-vanishing expectation value for a certain operator. 
For condensed matter model systems, possible applications of this effect have been studied for a variety of low-dimensional physical realizations, ranging from magnetic field induced
insulating behavior in semi-metallic structures~\cite{Khveshchenko:2001zza,Leal:2003sg} to anomalous Hall plateaus in graphene~\cite{Herbut:2007,Gusynin:2005pk,Gusynin:2006gn,Gusynin:2005iv,Gusynin:2006ym,Herbut:2008ui} 
and transport properties in $d$-wave superconductors~\cite{Semenoff:1998bk,Krishana:1999,Khveshchenko:2006}. It is interesting to note that magnetic catalysis can also go along with an induced anomalous magnetic moment of the fermionic excitations \cite{Ferrer:2008dy}.

The interplay between strong magnetic fields and chiral-symmetry breaking has
also become of topical relevance in particle physics. For instance,
non-central heavy-ion collisions go along with extreme magnetic fields \cite{Skokov:2009qp} that may even interfere with the topological structure of the QCD vacuum \cite{Kharzeev:2007jp,Fukushima:2008xe} or simply take a parametric influence on the chiral phases of QCD \cite{Shushpanov:1997sf,Cohen:2007bt,Zayakin:2008cy,Mizher:2010zb,Gatto:2010qs,Boomsma:2009yk,Skokov:2011ib}. Even though magnetic fields in strongly-interacting fermionic systems generically support chiral-condensate formation at least in mean-field-type approximations, further interactions such as gluonic back-reactions may also lead to inverse effects \cite{Bali:2011qj}. 

There exist already a few works in the literature on the large-$N_{\mathrm{f}}$ results for the Gross-Neveu model~\cite{Klimenko:1990rh,Klimenko:1991he,Krive:1992xh}.
Beyond leading-order results in a $1/N_{\mathrm{f}}$-expansion, magnetic catalysis was studied both with truncated Dyson-Schwinger equations (in $d=3$~\cite{Gusynin:1994xp,Alexandre:2000yf} and $d=4$~\cite{Leung:1996qy,Gusynin:1994xp,Gusynin:1999pq}) and Wilson-style
renormalization group (RG) equations (in $d=4$~\cite{Hong:1996pv,Hong:1997uw,Semenoff:1999xv}) in the context of fermionic models with short-ranged interactions and $\mathrm{QED}$. However, a clear RG picture in terms of a fixed-point analysis of the 
beta function appears to be still missing. Moreover, the inclusion of purely magnetically induced operators in the RG flow and their effect on the running coupling has not been considered so far. 
In this paper, we provide a clear renormalization mechanism for magnetic catalysis in the Gross-Neveu model. Similar observations are made in an RG study of magnetic catalysis in a QCD low-energy model \cite{Fukushima:2011aa}. Moreover, we identify terms that are compatible with the symmetries of the Gross-Neveu action in the presence of 
an external magnetic field. In principle, all operators compatible
with the symmetries of the initial action can be generated under RG
transformations. Therefore, our analysis addresses
the question as to whether the operator content of fermionic effective actions is modified in the IR by the very presence of such gauge backgrounds. 

The paper is organized as follows. In Sect.~\ref{sec:gn}, we define the three-dimensional, chirally symmetric Gross-Neveu model in an external magnetic field.
In Sect.~\ref{sec:gnsymm}, we collect the transformation rules under both discrete symmetry and chiral transformations, which we
will then employ in Sect.~\ref{sec:co} to study the Gross-Neveu theory space. Put differently, we give the transformation properties of 
all bilinear operators and list all quartic operators compatible with the Gross-Neveu symmetries. This includes the set of magnetically induced operators
(up to a specific mass dimension), the presence of which is completely sustained by the external magnetic field. To keep the presentation 
self-contained, in Sect.~\ref{sec:frg} we recapitulate the functional RG essentials for the one-particle irreducible (1PI) scheme. In Sect.~\ref{sec:beta} we
discuss the effect of the magnetic field on the behavior of the beta function for the fermionic coupling in the pointlike approximation in a qualitative manner. A quantitative discussion of the field dependence of the dynamically generated fermion mass inferred from the beta function is deferred to Sect.~\ref{sec:sg}. 
We conclude with Sect.~\ref{sec:cao} where we summarize our findings and give an outlook to our future work. Technical details and the derivation of the beta function are provided in
Appendices~\ref{App:sp}--\ref{App:tf}.  

\section{Gross-Neveu model in an external magnetic field}\label{sec:gn}

We consider the Gross-Neveu model~\cite{Gross:1974jv} within functional integral quantization. 
The microscopic degrees of freedom are represented by Grassmann-valued fields defined
over three-dimensional Euclidean space-time. For studying the effects of magnetic catalysis,
we minimally couple the Dirac fermions to a background gauge-potential that realizes the external
magnetic field. The defining local action to be quantized is given by
\begin{equation}
\label{eq:action} 
S[\barspinor,\spinor,\mathcal{A}]
=\int_{x}\mathcal{L}(\barspinor,\spinor,\partial_{\mu}\barspinor,\partial_{\mu}\spinor,\mathcal{A})
\end{equation}
with the Lagrangian density
\begin{equation}
\label{eq:lagrangian} 
\mathcal{L}
=\sum_{j=1}^{\Nf}\barspinor\mathrm{i}\fslash{D}[\mathcal{A}]\,\spinor
  + \sum_{i,j=1}^{\Nf}\altbarspinor\altspinor\frac{\bar{g}}{2
    \Nf}\barspinor\spinor,
\end{equation}
where $\int_{x}=\int d^{d} x$ is a shorthand for the integral over the
$d$-dimensional Euclidean space-time. Here, 
\begin{equation}
\fslash{D}[\mathcal{A}]=\gamma_{\mu}\left(\partial_{\mu}-\mathrm{i}q\mathcal{A}_{\mu}(x)\right),\quad\mu=0,1,2,
\end{equation}
denotes the covariant derivative acting on the spinor fields with $q$ the respective charge under the
Abelian $\mathrm{U}(1)$ gauge group. We consider $N_{\mathrm{f}}$ different flavor species indexed by $j=1,\dots,N_{\mathrm{f}}$, but keep the flavor number arbitrary.
This model enjoys a global $\mathrm{U}(N_{\mathrm{f}})$ flavor symmetry. The canonical mass
dimensions of the fields are given by $[\psi]=\frac{d-1}{2}$ and $[\mathcal{A}]=\frac{d-2}{2}$. This renders the gauge charge a dimensionful quantity with
$[q]=\frac{4-d}{2}$. As mentioned in Sect.~\ref{sec:intro}, we use a reducible four-component representation for the gamma matrices in this work, i.e.,
$d_{\gamma}=4$ where $d_{\gamma}=\mathrm{tr}\mathds{1}_{4}$ denotes the dimension of the representation space of the Dirac algebra and $\mathrm{tr}$ is the trace over spinor indices. Our choice for the $4\times 4$ representation of the Dirac algebra is the so-called chiral one. Here, the $\gamma_{5}$ matrix is in block-diagonal form. The matrices are explicitly given by
\be
\gamma_0=\tau_{2}\otimes\tau_{3}\,,\quad
\gamma_1=\tau_{2}\otimes\tau_{1}\,,\quad\gamma_{2}=\tau_{2}\otimes\tau_{2}\,
\ee
and
\be 
\gamma_{3}=\tau_{1}\otimes\tau_{0}\,,\quad
\gamma_{5}=\tau_{3}\otimes\tau_{0}\,,\quad\gamma_{35}\equiv\mathrm{i}\gamma_{3}\gamma_{5}
\ee
Here, the $\{\tau_i\}$'s denote the Pauli matrices which satisfy $\tau_i
\tau_j = \delta_{ij}\tau_0 + \mathrm{i}\epsilon_{ijk}\tau_k$, with
$i,j,k=1,2,3$ and $\tau_0=\mathds{1}_2$ is a $2\times 2$ unit matrix. The
gamma matrices satisfy the anti-commutation relations
\be
\{\gamma_\mu,\gamma_\nu\}=2\delta_{\mu\nu}\mathds{1}_{4}\,,
\ee 
where $\mu,\nu=0,1,2$. The matrices $\gamma_{3}$ and $\gamma_{5}$ anti-commute with all
$\gamma_{\mu}$, $\mu=0,1,2$. In this representation, $\gamma_{2}$, $\gamma_{3}$ and $\gamma_{5}$ are symmetric and real, while
$\gamma_{0}$, $\gamma_{1}$ and $\gamma_{35}$ are antisymmetric and purely imaginary $4\times 4$ matrices. One easily verifies
the following identities
\be 
\label{eq:idA}\gamma_{35}\gamma_{\mu}\gamma_{\nu} &=& \delta_{\mu\nu}\gamma_{35}+\mathrm{i}\epsilon_{\mu\nu\sigma}\gamma_{\sigma}, \\
\label{eq:idB}\mathrm{i}\gamma_{5}\gamma_{\mu}\gamma_{\nu} &=& \mathrm{i}\delta_{\mu\nu}\gamma_{5}+\mathrm{i}\epsilon_{\mu\nu\sigma}\gamma_{3}\gamma_{\sigma}, \\
\label{eq:idC}\mathrm{i}\gamma_{3}\gamma_{\mu}\gamma_{\nu} &=& \mathrm{i}\delta_{\mu\nu}\gamma_{3}-\mathrm{i}\epsilon_{\mu\nu\sigma}\gamma_{5}\gamma_{\sigma},
\ee
where the second and third lines follow from the first one by left or right multiplication with
an appropriate gamma matrix and making use of the Dirac algebra, and $\epsilon_{\mu\nu\sigma}$ is the completely
antisymmetric tensor.

We chose the vector potential in the gauge
\begin{equation}
\mathcal{A}(x)=\left(0,0,x_{1} B\right)^{T}.
\end{equation}
Within the three-dimensional formulation, the magnetic field appears as a 
pseudo-scalar quantity. This corresponds to the fact that the magnetic flux
associated to a magnetic field that is aligned perpendicularly to the $(x_{1},x_{2})$-plane
can penetrate this plane with two different orientations.  

At zero magnetic field, $B\equiv 0$, the model depends on a single parameter~\cite{Braun:2010tt}, namely, the dimensionful coupling constant $\bar{g}$ with mass dimension $2-d$. In the sense of
statistical physics, the coupling appears to correspond to an irrelevant operator
within the perturbative Gaussian classification. However, this naive scaling analysis
does not yield the correct picture for the infrared behavior of this model, as we will explain in detail in Sect.~\ref{sec:beta}. 
From the point of view of quantum field theory, the Gross-Neveu model provides an example of
a nonperturbatively renormalizable field theory~\cite{Gawedzki:1985jn}, and the $(\bar\psi\psi)^2$ coupling becomes RG relevant near a non-Gaussian fixed point \cite{Braun:2010tt}.

\section{Gross-Neveu symmetries}\label{sec:gnsymm}

While the construction of local operators that conform to Lorentz symmetry can
be ensured by taking products of operators and performing suitable contractions over 
tensor indices, the transformation properties under discrete space-time symmetries require more work.

In Ref.~\cite{Gies:2010st} explicit representations for discrete space-time transformations as realized on 4-component
spinor fields over three dimensional Euclidean space were given as follows:
\be
\label{eq:cc}\mathcal{C}:\,C_{\xi}  &=& \frac{1}{2}\left[(1+\xi)\gamma_{2}\gamma_{3}+\mathrm{i}(1-\xi)\gamma_{2}\gamma_{5}\right],\\
\label{eq:pt}\mathcal{P}:\,P_{\zeta}&=& \frac{1}{2}\left[(1+\zeta)\gamma_{1}\gamma_{3}+\mathrm{i}(1-\zeta)\gamma_{1}\gamma_{5}\right],\\
\label{eq:tr}\mathcal{T}:\,T_{\eta} &=& \frac{1}{2}\left[(1+\eta)\gamma_{1}+\mathrm{i}(1-\eta)\gamma_{2}\gamma_{0}\right],
\ee
where $\mathcal{C}$, $\mathcal{P}$, $\mathcal{T}$ denote charge conjugation, parity inversion and time
reversal. To each transformation, there exists an associated unitary $4\times 4$ matrix, which we denote by $C_{\xi}$, $P_{\zeta}$ and $T_{\eta}$, respectively.
Other possible conventions were given for example in~\cite{Gomes:1990ed}. However, we will stick to the definition
as displayed in Eqs.~\eqref{eq:cc},~\eqref{eq:pt} and ~\eqref{eq:tr}. Concerning parity, it is worthwhile to mention that in three space-time dimensions, parity inversion is properly defined
by $(x_{0},x_{1},x_{2})\mapsto(x_{0},-x_{1},x_{2})$, i.e. only one of the spatial components is reversed. This is due to the fact, that
in our case, only a single generator for rotations exists. The above definition ensures that parity inversion is not an element of the
connected component of rotations containing the identity.
As can be seen in Eqs.~\eqref{eq:cc}-\eqref{eq:tr}, there exists an entire family of realizations of discrete transformations depending on pure phase variables $\xi$, $\zeta$ and $\eta$ with
unit modulus. We will simply set $\xi=\zeta=\eta=1$ in the following and omit the subscripts on the transformation matrices. The realization of space-time transformations
on spinor fields reads for charge conjugation
\be\label{eq:cspinor} 
\mathcal{C}\psi(x)\mathcal{C}^{-1} &=& (\psibar C)^{T},\,\,\mathcal{C}\psibar(x)\mathcal{C}^{-1} = -(C^{\dagger} \psi(x))^{T},
\ee
parity
\be\label{eq:pspinor} 
\mathcal{P}\psi(x)\mathcal{P}^{-1} &=& P\psi(\tilde{x}),\,\,\mathcal{P}\psibar(\tilde{x})\mathcal{P}^{-1} = \psibar(\tilde{x}) P^{\dagger},
\ee
and time-reversal
\be\label{eq:tspinor} 
\mathcal{T}\psi(x)\mathcal{T}^{-1} &=& T \psi(\hat{x}),\,\,\mathcal{T}\psibar(x)\mathcal{T}^{-1} = \psibar(\hat{x})T^{\dagger}.
\ee
Here we follow the notation as given in~\cite{Gomes:1990ed}, and define
\be 
\tilde{x}=(x_{0},-x_{1},x_{2})^{T},\quad\hat{x}=(-x_{0},x_{1},x_{2})^{T}.
\ee
Note that by virtue of $\mathcal{T}\,\mathrm{i}\,\mathcal{T}^{-1}=-\mathrm{i}$, time-reversal is an anti-unitary transformation.
We will consider a theory to be symmetric under $\mathcal{C}$-, $\mathcal{P}$- and $\mathcal{T}$-transformations,
if its Lagrangian density obeys
\be 
\mathcal{C}\mathcal{L}(x)\mathcal{C}^{-1} &=& \mathcal{L}(x), \\
\mathcal{P}\mathcal{L}(x)\mathcal{P}^{-1} &=& \mathcal{L}(\tilde{x}), \\
\mathcal{T}\mathcal{L}(x)\mathcal{T}^{-1} &=& \mathcal{L}(\hat{x}),
\ee
such that a simultaneous transformation acting on fields \emph{and} coordinates leaves $\mathcal{L}$ invariant
as a function of space-time coordinates.

The Gross-Neveu model as defined in \Eqref{eq:lagrangian} is symmetric under the \emph{discrete} chiral transformation
\be\label{eq:chiral}
\psi\mapsto\gamma_{5}\psi,\quad\psibar\mapsto-\psibar\gamma_{5}.
\ee
Since the composite operator $\psibar\psi$ transforms into $-\psibar\psi$ under this discrete
chiral transformation, a finite expectation value $\langle\psibar\psi\rangle\neq 0$ in a given quantum state signals the 
breakdown of chiral symmetry. In the following, we denote the discrete symmetry group by $\mathds{Z}_{2}^{5}=\{\mathds{1},\gamma_{5}\}$. 
However, there exists also a continuous Abelian chiral symmetry, generated
by $\gamma_{35}$:
\be\label{eq:conti_chiral}
\psi\mapsto\mathrm{e}^{\mathrm{i}\varphi\gamma_{35}}\psi,\quad\psibar\mapsto\psibar\mathrm{e}^{-\mathrm{i}\varphi\gamma_{35}}.
\ee
It is easy to see that the element $\mathrm{e}^{\mathrm{i}\frac{\pi}{2}\gamma_{35}}\in\mathrm{U}^{35}(1)$ combined with the non-trivial $\mathds{Z}_{2}^{3}$ transformation
\be 
\psi\mapsto\gamma_{3}\psi,\quad\psibar\mapsto-\psibar\gamma_{3}
\ee
leads us back to the transformation \Eqref{eq:chiral}. In this respect, the discrete $\mathds{Z}_{2}^{3}$ does not yield a new symmetry of the theory and
we will henceforth choose $\mathds{Z}_{2}^{5}$ and $\mathrm{U}^{35}(1)$ to define the chiral symmetries of 
the three dimensional Gross-Neveu model. It is perhaps worth mentioning that for the free theory, i.e. $\bar{g}=0$, the symmetry
transformations act independently on all flavor species $j$, $j=1,\dots,N_{\mathrm{f}}$. For finite couplings, however, all flavors are subject 
to a simultaneous chiral $\mathds{Z}_{2}^{5}$ transformation, as we expect from the symmetry-breaking pattern induced by the interaction term.

To conclude this section, we summarize the properties of electromagnetic quantities under discrete space-time transformations.
The gauge potential $\mathcal{A}$ transforms as
\be 
\mathcal{C}\mathcal{A}_{\mu}(x)\mathcal{C}^{-1}&=&-\mathcal{A}_{\mu}(x),\\
\label{eq:Apt}\mathcal{P}\mathcal{A}_{\mu}(x)\mathcal{P}^{-1}&=&\tilde{\mathcal{A}}_{\mu}(\tilde{x}),\\
\label{eq:Atr}\mathcal{T}\mathcal{A}_{\mu}(x)\mathcal{T}^{-1}&=&-\hat{\mathcal{A}}_{\mu}(\hat{x}),
\ee
where $\tilde{\mathcal{A}}=(\mathcal{A}_{0},-\mathcal{A}_{1},\mathcal{A}_{2})^{T}$ and $\hat{\mathcal{A}}=(-\mathcal{A}_{0},\mathcal{A}_{1},\mathcal{A}_{2})^{T}$.
The field-strength tensor $\mathcal{F}_{\mu\nu}=\partial_{\mu}\mathcal{A}_{\nu}-\partial_{\nu}\mathcal{A}_{\mu}$ accordingly obeys
\be 
\mathcal{C}\mathcal{F}_{\mu\nu}(x)\mathcal{C}^{-1}&=&-\mathcal{F}_{\mu\nu}(x),\\
\label{eq:FSpt}\mathcal{P}\mathcal{F}_{\mu\nu}(x)\mathcal{P}^{-1}&=&\tilde{\mathcal{F}}_{\mu\nu}(\tilde{x}),\\
\label{eq:FStr}\mathcal{T}\mathcal{F}_{\mu\nu}(x)\mathcal{T}^{-1}&=&-\hat{\mathcal{F}}_{\mu\nu}(\hat{x}).
\ee
Here, by $\tilde{\mathcal{F}}_{\mu\nu}$ and $\hat{\mathcal{F}}_{\mu\nu}$ we denote matrices which result from plugging Eqs.~\eqref{eq:Apt} and \eqref{eq:Atr} into
the definition of the field-strength tensor. The so-called dual `field-strength' $F_{\mu}\equiv\frac{1}{2}\epsilon_{\mu\nu\sigma}\mathcal{F}_{\nu\sigma}$ 
becomes a pseudo-vector quantity in three dimensions.
Thus, it behaves as
\be 
\mathcal{C}F_{\mu}(x)\mathcal{C}^{-1}&=&-F_{\mu}(x),\\
\mathcal{P}F_{\mu}(x)\mathcal{P}^{-1}&=&-\tilde{F}_{\mu}(\tilde{x}),\\
\mathcal{T}F_{\mu}(x)\mathcal{T}^{-1}&=&\hat{F}_{\mu}(\hat{x}).
\ee
The simplest Lorentz, gauge and $\mathcal{C}$-, $\mathcal{P}$-, $\mathcal{T}$-invariants that can be built
from the field strength and its dual are
\be 
\mathcal{F}_{\mu\nu}^{2}\quad\text{and}\quad F_{\mu}^{2}.
\ee
Since $F_{\mu}^{2}=\frac{1}{2}\mathcal{F}_{\mu\nu}^{2}$ there is only one linearly independent invariant.

%

\section{Classification of compatible operators}\label{sec:co}

Having collected the prerequisites to study all operators that are in principle compatible
symmetry-wise with the Lagrangian density \Eqref{eq:lagrangian}, we will give an exhaustive classification 
on the level of fermionic bilinears in Sect.~\ref{subsec:bo} and then proceed to quartic fermionic terms including
purely magnetically induced operators in Sect.~\ref{subsec:qo}. 

\subsection{Bilinear Operators}\label{subsec:bo}

Fermionic bilinears are the building blocks of an action for fermionic degrees of freedom. Terms entering
the quadratic part of the action define the inverse bare propagator of the theory. 
In our case, we need to contract both spinor and Lorentz indices to form an appropriate scalar quantity. But, due
to the presence of the gauge background, we can as well use the gauge-invariant field-strength tensor and its dual to build Lorentz-invariant fermion bilinears. 
We consider only those bilinears which are Lorentz symmetric as a building block of the effective action. By use of Eqs.~\eqref{eq:cspinor}--\eqref{eq:tspinor} and Eqs.~\eqref{eq:chiral} and~\eqref{eq:conti_chiral} we obtain the behavior of a given bilinear under discrete space-time and chiral transformations. The results are collected in Tables~\ref{tab1},~\ref{tab2}, and~\ref{tab3}. For turning operators with vector or tensor structure into 
Lorentz-invariant scalars, we contract as indicated above with $F_{\mu}$ or $\mathcal{F}_{\mu\nu}$. By virtue of the canonical mass dimension of the fermion fields $[\psi]=\frac{d-1}{2}$, the
mass dimension of a (non-derivative) fermion bilinear is $[\psibar\Gamma_{O}\psi]=d-1$ with $\Gamma_{S}\in\{\mathds{1},
\gamma_{3},\gamma_{5}, \gamma_{35}\}$ (scalar/pseudo-scalar), $\Gamma_{V}\in\{\gamma_{\mu},\gamma_{3}\gamma_{\mu},\gamma_{5}\gamma_{\mu},\gamma_{35}\gamma_{\mu}$ (vector/axial vector), or $\Gamma_{T}\in\{\sigma_{\mu\nu},\gamma_{3}\sigma_{\mu\nu},\gamma_{5}\sigma_{\mu\nu},\gamma_{35}\sigma_{\mu\nu}\}$ (tensor/pseudotensor). Here,
\be 
\sigma_{\mu\nu}=\frac{\mathrm{i}}{2}[\gamma_{\mu},\gamma_{\nu}],\quad\mu,\nu=0,1,2,
\ee
is the set of generators for (Euclidean) Lorentz transformations. For the contracted terms
we obtain $[q F_{\mu}(\psibar\Gamma_{\mu}\psi)]=d+1$ and $[q \mathcal{F}_{\mu\nu}(\psibar\Gamma_{\mu\nu}\psi)]=d+1$, since
$[F_{\mu}]=[\mathcal{F}_{\mu\nu}]=\frac{d}{2}$ and $[q]=\frac{4-d}{2}$. Taking into account that the space-time integral in a local action contributes $[\int_{x}]=-d$ to the total
mass dimension of a given local operator, these operators naively correspond to irrelevant directions in theory space. The charge $q$ needs to be included in this
consideration, since we are interested in magnetically \emph{induced} phenomena. We could also perform two contractions of $q F_{\mu}$ with an appropriate tensor structure. This would inevitably increase the mass dimension by 2 and render this term even more power-counting irrelevant compared to mass terms (with mass dimension $-1$) and kinetic operators with mass dimension $0$. However, from Tables~\ref{tab1}--\ref{tab3}, we see that none of these bilinear operators containing the magnetic field are compatible with the symmetries of the Gross-Neveu model. They cannot be generated during the RG flow in a continuous fashion. 
Combinations like $\mathcal{F}_{\mu\nu}(\psibar\sigma_{\mu\nu}\psi)$ and $F_{\mu}(\psibar\gamma_{35}\gamma_{\mu}\psi)$ conform 
to space-time symmetries, but violate the chiral symmetries of the Gross-Neveu action. Chirally invariant combinations containing two $F_{\mu}$ 
fields are not even symmetric with respect to $\mathcal{C}$, $\mathcal{P}$ and $\mathcal{T}$.

To conclude, we would like to comment that each $F_{\mu}$ contraction with an appropriate vector bilinear can be rewritten as a contraction of
$\mathcal{F}_{\mu\nu}$ with a tensor bilinear by means of Eqs.~\eqref{eq:idA}-\eqref{eq:idC}:
\be 
\label{eq:idVTA}F_{\mu}(\psibar\gamma_{\mu}\psi) & = & -\frac{1}{2}\mathcal{F}_{\mu\nu}(\psibar\gamma_{35}\sigma_{\mu\nu}\psi),\\
F_{\mu}(\psibar\gamma_{3}\gamma_{\mu}\psi) & = & -\frac{\mathrm{i}}{2}\mathcal{F}_{\mu\nu}(\psibar\gamma_{5}\sigma_{\mu\nu}\psi),\\
F_{\mu}(\psibar\gamma_{5}\gamma_{\mu}\psi) & = & +\frac{\mathrm{i}}{2}\mathcal{F}_{\mu\nu}(\psibar\gamma_{3}\sigma_{\mu\nu}\psi),\\
\label{eq:idVTD}F_{\mu}(\psibar\gamma_{35}\gamma_{\mu}\psi) & = & -\frac{1}{2}\mathcal{F}_{\mu\nu}(\psibar\sigma_{\mu\nu}\psi).
\ee
%

\begin{table*}[h!t]\center
\begin{tabular}{p{70pt}||p{70pt}|p{70pt}|p{70pt}|p{70pt}|p{70pt}p{0pt}}
\centering  & \centering $\mathcal{C}$ & \centering $\mathcal{P}$ & \centering $\mathcal{T}$ & \centering $\mathds{Z}_{2}^{5}$ & \centering $\mathrm{U}^{35}(1)$ & \\ \hline\hline
\centering $(\psibar\psi)(x)$ & \centering $(\psibar\psi)(x)$ & \centering $(\psibar\psi)(\tilde{x})$ & \centering $(\psibar\psi)(\hat{x})$ & \centering $-(\psibar\psi)(x)$ & \centering $(\psibar\psi)(x)$ & \\ \hline
\centering $(\psibar\gamma_{3}\psi)(x)$ & \centering $-(\psibar\gamma_{3}\psi)(x)$ & \centering $-(\psibar\gamma_{3}\psi)(\tilde{x})$ & \centering $-(\psibar\gamma_{3}\psi)(\hat{x})$ & \centering $(\psibar\gamma_{3}\psi)(x)$ & \centering $(\psibar\gamma_{3}\mathrm{e}^{2\mathrm{i}\varphi\gamma_{35}}\psi)(x)$ & \\ \hline
\centering $(\psibar\gamma_{5}\psi)(x)$ & \centering $(\psibar\gamma_{5}\psi)(x)$ & \centering $(\psibar\gamma_{5}\psi)(\tilde{x})$ & \centering $-(\psibar\gamma_{5}\psi)(\hat{x})$ & \centering $-(\psibar\gamma_{5}\psi)(x)$ & \centering $(\psibar\gamma_{5}\mathrm{e}^{2\mathrm{i}\varphi\gamma_{35}}\psi)(x)$ & \\ \hline
\centering $(\psibar\gamma_{35}\psi)(x)$ & \centering $(\psibar\gamma_{35}\psi)(x)$ & \centering $-(\psibar\gamma_{35}\psi)(\tilde{x})$ & \centering $-(\psibar\gamma_{35}\psi)(\hat{x})$ & \centering $(\psibar\gamma_{35}\psi)(x)$ & \centering $(\psibar\gamma_{35}\psi)(x)$ & \\
\end{tabular}
\caption{Classification of scalar/pseudo-scalar fermion bilinears $(\psibar\Gamma_{S}\psi)(x)$ according to their behavior under discrete space-time and chiral transformations.}
\label{tab1}
\end{table*}

\begin{table*}[h!t]\center
\begin{tabular}{p{70pt}||p{70pt}|p{70pt}|p{70pt}|p{70pt}|p{70pt}p{0pt}}
\centering  & \centering $\mathcal{C}$ & \centering $\mathcal{P}$ & \centering $\mathcal{T}$ & \centering $\mathds{Z}_{2}^{5}$ & \centering $\mathrm{U}^{35}(1)$ & \\ \hline\hline
\centering $(\psibar\gamma_{\mu}\psi)(x)$ & \centering $-(\psibar\gamma_{\mu}\psi)(x)$ & \centering $(\psibar\tilde{\gamma}_{\mu}\psi)(\tilde{x})$ & \centering $-(\psibar\hat{\gamma}_{\mu}\psi)(\hat{x})$ & \centering $(\psibar\gamma_{\mu}\psi)(x)$ & \centering $(\psibar\gamma_{\mu}\psi)(x)$ & \\ \hline
\centering $(\psibar\gamma_{3}\gamma_{\mu}\psi)(x)$ & \centering $-(\psibar\gamma_{3}\gamma_{\mu}\psi)(x)$ & \centering $-(\psibar\gamma_{3}\tilde{\gamma}_{\mu}\psi)(\tilde{x})$ & \centering $(\psibar\gamma_{3}\hat{\gamma}_{\mu}\psi)(\hat{x})$ & \centering $-(\psibar\gamma_{3}\gamma_{\mu}\psi)(x)$ & \centering $(\psibar\gamma_{3}\gamma_{\mu}\mathrm{e}^{2\mathrm{i}\varphi\gamma_{35}}\psi)(x)$ & \\ \hline
\centering $(\psibar\gamma_{5}\gamma_{\mu}\psi)(x)$ & \centering $(\psibar\gamma_{5}\gamma_{\mu}\psi)(x)$ & \centering $(\psibar\gamma_{5}\tilde{\gamma}_{\mu}\psi)(\tilde{x})$ & \centering $(\psibar\gamma_{5}\hat{\gamma}_{\mu}\psi)(\hat{x})$ & \centering $(\psibar\gamma_{5}\gamma_{\mu}\psi)(x)$ & \centering $(\psibar\gamma_{5}\gamma_{\mu}\mathrm{e}^{2\mathrm{i}\varphi\gamma_{35}}\psi)(x)$ & \\ \hline
\centering $(\psibar\gamma_{35}\gamma_{\mu}\psi)(x)$ & \centering $-(\psibar\gamma_{35}\gamma_{\mu}\psi)(x)$ & \centering $-(\psibar\gamma_{35}\tilde{\gamma}_{\mu}\psi)(\tilde{x})$ & \centering $(\psibar\gamma_{35}\hat{\gamma}_{\mu}\psi)(\hat{x})$ & \centering $-(\psibar\gamma_{35}\gamma_{\mu}\psi)(x)$ & \centering $(\psibar\gamma_{35}\gamma_{\mu}\psi)(x)$ & \\
\end{tabular}
\caption{Classification of vector/axial-vector fermion bilinears
  $(\psibar\Gamma_{V}\psi)(x)$ according to their behavior under discrete
  space-time and chiral transformations. Here, we have defined $\tilde{\gamma}=(\gamma_{0},
-\gamma_{1},\gamma_{2})^{T}$ and $\hat{\gamma}=(-\gamma_{0},\gamma_{1},\gamma_{2})^{T}$. The bilinear $\psibar\gamma_{35}\gamma_{\mu}\psi$ can be shown to be related to $\psibar\sigma_{\mu\nu}\psi$, cf. \Eqref{eq:idA}.}
\label{tab2}
\end{table*}

\begin{table*}[h!t]\center
\begin{tabular}{p{70pt}||p{70pt}|p{70pt}|p{70pt}|p{70pt}|p{70pt}p{0pt}}
\centering  & \centering $\mathcal{C}$ & \centering $\mathcal{P}$ & \centering $\mathcal{T}$ & \centering $\mathds{Z}_{2}^{5}$ & \centering $\mathrm{U}^{35}(1)$ & \\ \hline\hline
\centering $(\psibar\sigma_{\mu\nu}\psi)(x)$ & \centering $-(\psibar\sigma_{\mu\nu}\psi)(x)$ & \centering $(\psibar\tilde{\sigma}_{\mu\nu}\psi)(\tilde{x})$ & \centering $(\psibar\hat{\sigma}_{\mu\nu}\psi)(\hat{x})$ & \centering $-(\psibar\sigma_{\mu\nu}\psi)(x)$ & \centering $(\psibar\sigma_{\mu\nu}\psi)(x)$ & \\ \hline
\centering $(\psibar\gamma_{3}\sigma_{\mu\nu}\psi)(x)$ & \centering $(\psibar\gamma_{3}\sigma_{\mu\nu}\psi)(x)$ & \centering $-(\psibar\gamma_{3}\tilde{\sigma}_{\mu\nu}\psi)(\tilde{x})$ & \centering $(\psibar\gamma_{3}\hat{\sigma}_{\mu\nu}\psi)(\hat{x})$ & \centering $(\psibar\gamma_{3}\sigma_{\mu\nu}\psi)(x)$ & \centering $(\psibar\gamma_{3}\sigma_{\mu\nu}\mathrm{e}^{2\mathrm{i}\varphi\gamma_{35}}\psi)(x)$ & \\ \hline
\centering $(\psibar\gamma_{5}\sigma_{\mu\nu}\psi)(x)$ & \centering $-(\psibar\gamma_{5}\sigma_{\mu\nu}\psi)(x)$ & \centering $(\psibar\gamma_{5}\tilde{\sigma}_{\mu\nu}\psi)(\tilde{x})$ & \centering $(\psibar\gamma_{5}\hat{\sigma}_{\mu\nu}\psi)(x)$ & \centering $-(\psibar\gamma_{5}\sigma_{\mu\nu}\psi)(x)$ & \centering $(\psibar\gamma_{5}\sigma_{\mu\nu}\mathrm{e}^{2\mathrm{i}\varphi\gamma_{35}}\psi)(x)$ & \\ \hline
\centering $(\psibar\gamma_{35}\sigma_{\mu\nu}\psi)(x)$ & \centering $-(\psibar\gamma_{35}\sigma_{\mu\nu}\psi)(x)$ & \centering $-(\psibar\gamma_{35}\tilde{\sigma}_{\mu\nu}\psi)(\tilde{x})$ & \centering $(\psibar\gamma_{35}\hat{\sigma}_{\mu\nu}\psi)(\hat{x})$ & \centering $(\psibar\gamma_{35}\sigma_{\mu\nu}\psi)(x)$ & \centering $(\psibar\gamma_{35}\sigma_{\mu\nu}\psi)(x)$ & \\
\end{tabular}
\caption{Classification of tensor/pseudo-tensor fermion bilinears $(\psibar\Gamma_{T}\psi)(x)$ according to their behavior under discrete space-time and chiral transformations. The
 matrices $\tilde{\sigma}_{\mu\nu}$ and $\hat{\sigma}_{\mu\nu}$ carry the same sign structure as $\tilde{\mathcal{F}}_{\mu\nu}$ and $\hat{\mathcal{F}}_{\mu\nu}$ in \Eqref{eq:FSpt} and~\eqref{eq:FStr}, respectively.
 The bilinear $\psibar\gamma_{35}\sigma_{\mu\nu}\psi$ can be shown to be equivalent to $\psibar\gamma_{\mu}\psi$, cf. \Eqref{eq:idC}. It also holds that $\psibar\gamma_{3/5}\sigma_{\mu\nu}\psi\sim\psibar\gamma_{5/3}\gamma_{\mu}\psi$.}
\label{tab3}
\end{table*}

\subsection{Quartic Operators}\label{subsec:qo}

Quartic fermionic terms describe the two-body interaction processes in our theory. We first
ask for terms 
\be\label{eq:quartic}
\sum_{i,j=1}^{\Nf}(\altbarspinor\Gamma_{X}\altspinor)(\barspinor\Gamma_{Y}\spinor)
\ee 
which obey the symmetries of the original Gross-Neveu action and have canonical mass dimension
$[(\altbarspinor\Gamma_{X}\altspinor)(\barspinor\Gamma_{Y}\spinor)]=2d-2$. In a next step, we analyze contributions with mass dimension $2d$,
i.e., quartic terms that include a contraction with a mass dimension $2$ object. In principle, one could also move to operators with higher mass dimension. But with increasing mass dimension our naive expectation is that these operators become increasingly irrelevant for the IR behavior of the theory, even near a non-Gaussian fixed point. 

Note that terms as captured by \Eqref{eq:quartic} are composed of flavor singlets. Invariant sums of bilinears with
off-diagonal flavor structure can be brought into this singlet-singlet form by an appropriate 
Fierz transformation. The result is in general a sum over several singlet-singlet contributions
with different Lorentz structure.   

Since a discrete chiral symmetry is less restrictive than continuous chiral symmetry, the number of allowed quartic terms appears to be quite large. However, the requirement for invariance under the \emph{continuous} $\mathrm{U}^{35}(1)$ symmetry remedies the situation somewhat. The allowed terms in the absence of a magnetic field are exhausted by $(\psibar\psi)^{2}$, $(\psibar\gamma_{\mu}\psi)^{2}$, $(\psibar\sigma_{\mu\nu}\psi)^{2}$ and 
$(\psibar\gamma_{35}\psi)^{2}$, $(\psibar\gamma_{35}\gamma_{\mu}\psi)^{2}$,
$(\psibar\gamma_{35}\sigma_{\mu\nu}\psi)^{2}$ as well as the invariant combinations $(\psibar\gamma_{3}\psi)^{2}+(\psibar\gamma_{5}\psi)^{2}$, $(\psibar\gamma_{3}\gamma_{\mu}\psi)^{2}+(\psibar\gamma_{5}\gamma_{\mu}\psi)^{2}$ where we have omitted all flavor indices.
Actually $(\psibar\gamma_{35}\gamma_{\mu}\psi)^{2}$ and $(\psibar\gamma_{35}\sigma_{\mu\nu}\psi)^{2}$ can be written as
\be 
(\psibar\gamma_{35}\sigma_{\mu\nu}\psi)^{2} & = & \frac{1}{2}(\psibar\gamma_{\mu}\psi)^{2},\\
(\psibar\gamma_{35}\gamma_{\mu}\psi)^{2} & = & \frac{1}{2}(\psibar\sigma_{\mu\nu}\psi)^{2},
\ee 
since they are simply linear combinations of the Dirac algebra basis elements.  
The remaining six quartic terms are mutually independent.  
%
%
%
They parametrize the `Gross-Neveu theory space' without an external magnetic field in the pointlike limit. It is interesting to note that an RG flow calculation in the pointlike limit does actually not make use of this full theory space, but only reproduces the Gross-Neveu coupling in the derivative expansion. This might be accidental or could point to a further hidden symmetry.

We now consider quartic fermionic terms that can be built by contracting once with
$q \mathcal{F}_{\mu\nu}$ or $q F_{\mu}$. Due to the identities Eqs.~\eqref{eq:idVTA}-\eqref{eq:idVTD}, it is sufficient to
consider contractions with $q F_{\mu}$ only. From Tables~\ref{tab1}-\ref{tab3} we conclude, that the following terms are allowed
\be\label{eq:scalar_tensor} 
(\psibar\psi)(\psibar\gamma_{35}\gamma_{\mu}F_{\mu}\psi)=-\frac{1}{2}(\psibar\psi)(\psibar\sigma_{\mu\nu}\mathcal{F}_{\mu\nu}\psi),
\ee 
\be 
(\psibar\gamma_{35}\psi)(\psibar\gamma_{\mu}F_{\mu}\psi),
\ee 
and
\be 
(\psibar\gamma_{3}\psi)(\psibar\gamma_{5}\gamma_{\mu}F_{\mu}\psi)+(\psibar\gamma_{5}\psi)(\psibar\gamma_{3}\gamma_{\mu}F_{\mu}\psi).
\ee 
Of course, a $\mathrm{U}^{35}(1)$ invariant operator with fully contracted Lorentz indices squared is also compatible with the Gross-Neveu symmetries, but
has mass dimension larger or equal to $2d+2$. One could also consider functions of the invariant $ (q F_{\mu})^{2}$ times a bilinear squared. This, however, corresponds to operators with arbitrarily high mass dimension. In
Appendix~\ref{App:rg} we show that starting from the naive Gross-Neveu action \Eqref{eq:action}, a term of the type \Eqref{eq:scalar_tensor}
is indeed generated in an infinitesimal RG step. In phases with broken chiral symmetry, such an operator may give rise to an anomalous magnetic moment, as observed in~\cite{Ferrer:2008dy}.

\section{Functional RG equation}\label{sec:frg}

In this section we briefly summarize the essentials of the (functional) renormalization group. In the space of
couplings $g_i$ that parametrize a given action, the RG provides a vector field $\boldsymbol
\beta$, summarizing the RG $\beta$ functions for these couplings $(\boldsymbol\beta)_i
= \beta_{g_i}(g_1,g_2,\dots)\equiv \pat g_i$. As the full content of a quantum
theory can be specified in terms of generating functionals for correlation
functions or vertices, we can more generally study the RG behavior of a generating
functional. Introducing an IR-regulated effective average action $\Gamma_k$ by
inserting a scale-dependent regulator into the Gaussian measure of the functional integral representing
the generating functional of vacuum correlators, $Z_{k}[J]$, the RG flow of this action is determined by the Wetterich equation \cite{Wetterich:1992yh}
\begin{equation}\label{flowequation}
    \partial_t\Gamma_k[\Phi]
=\frac{1}{2}\mathrm{STr}\left\{\left[\Gamma^{(2)}_k[\Phi]+R_k\right]^{-1}(\partial_tR_k)\right\},
        \;\, \pat=k\frac{d}{dk}      .
\end{equation}
More precisely, $\Gamma_{k}[\Phi]$ is the modified Legendre transform
\be
\Gamma_k[\Phi]=(J,\Phi)-W_{k}[J]-\frac{1}{2}(\Phi,R_{k}\Phi)
\ee
of the scale-dependent generating functional
\be 
W_{k}[J]=\log Z_{k}[J]
\ee
for connected correlators and $\Gamma^{(2)}_k$ is the second functional derivative with respect to the
field $\Phi$, representing an appropriate collection of field variables for all bosonic or
fermionic degrees of freedom. The brackets $(\,\cdot\,,\,\cdot\,)$ are shorthand for suitable contractions of generalized indices.
The function $R_k$ denotes a momentum-dependent regulator that suppresses IR modes below a momentum scale $k$. $\mathrm{STr}$ denotes the so-called
super-trace operation, which simply takes into account the minus sign from the loop contributions from
the fermionic sector. The solution to the Wetterich equation provides for an RG trajectory in the space of all
action functionals, also known as {\em theory space}. The trajectory interpolates between
the bare action $S_\Lambda$ to be quantized $\Gamma_{k\to\Lambda}\to
S_\Lambda$ and the full quantum effective action $\Gamma=\Gamma_{k\to 0}$,
being the generating functional of 1PI correlation functions; for reviews with a focus on fermionic systems, see
\cite{Berges:2000ew,Pawlowski:2005xe,Gies:2006wv,Metzner:2011cw} and particularly \cite{Braun:2011pp}.
Parametrizing the effective average action $\Gamma_k$ by a
set of generalized dimensionless couplings $g_i$, the
Wetterich equation provides us with the corresponding RG flow $\pat g_i=
\beta_{g_i}(g_1,g_2,\dots) $. A fixed point $g_{i,\ast}$ is defined by
\begin{equation}
 \beta_i(g_{1,{\ast}},g_{2,{\ast}},...)=0\ \forall \ i\,.
\end{equation}
In general, this map of a quantum field theory to a flow in coupling space allows
to extract universal physical information from the fixed points and
from the associated manifolds 
controlling the flow toward the infrared.
%
\section{Magnetic beta function}\label{sec:beta}

In the following, we will elaborate on how the phenomenon of magnetic catalysis manifests 
itself in the IR behavior of the beta function for the coupling $\bar{g}$.  
The beta function for vanishing gauge background was derived previously within the formalism of the functional
renormalization group~\cite{Braun:2010tt}. Explicitly, it is given by
\be
\beta_{g}\equiv\partial_t g = (d - 2)g - 4 \left(\frac{N_{\mathrm{f}}\,d_{\gamma}-2}{N_{\mathrm{f}}}\right) v_d\,l_1^{F}(0)\, g^2\,
\label{eq:fourfermionflow}
\ee
with the dimensionless coupling
\be
g = k^{d-2} \bar{g}.
\ee
In Eq.~\eqref{eq:fourfermionflow}, $d_{\gamma}=\mathrm{tr}\mathds{1}_{4}=4$ for the present reducible
 fermion representation, and $v_{d}^{-1}=2^{d+1}\pi^{d/2}\Gamma(d/2)$. For the definition of the
threshold function $l_1^{F}(\omega)$ parametrizing the RG-scheme
dependence, see Appendix~\ref{App:tf}. Aside from the Gaussian fixed
point of a noninteracting theory, this beta function has also a
non-Gaussian, i.e., strong-coupling fixed point.  At this non-Gaussian fixed point the scaling dimension is altered from
$\Theta_{\mathrm{Gau\text{ss}}}(\bar{g})<0$ at the Gaussian one to
$\Theta_{\mathrm{nonGau\text{ss}}}(\bar{g})>0$,
cf.~\cite{Braun:2010tt}. So in contrast to a naive scaling analysis,
the coupling actually corresponds to a relevant direction in theory
space at this non-trivial zero of the beta function.  This fixed-point
structure is related to the known nonperturbative renormalizability of
the Gross-Neveu model.  The renormalization program was performed
within a $1/N_{\mathrm{f}}$-expansion ~\cite{Gawedzki:1985jn}, and it
was argued that the theory is renormalizable to all orders in this
expansion.  With the help of the renormalization group, the theory was
shown to be asymptotically safe, i.e., the RG flow contains
trajectories that eventually hit the non-trivial fixed point with
finite dimensionless couplings if the scale $k$ (and implicitly the UV
cutoff $\Lambda$) is sent to infinity~\cite{Braun:2010tt}.
\begin{figure}
\includegraphics[scale=0.15]{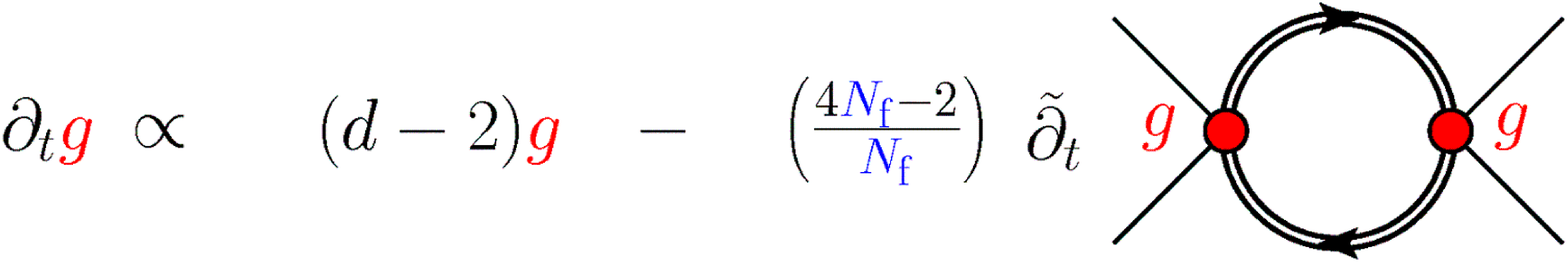}
\includegraphics[scale=0.15]{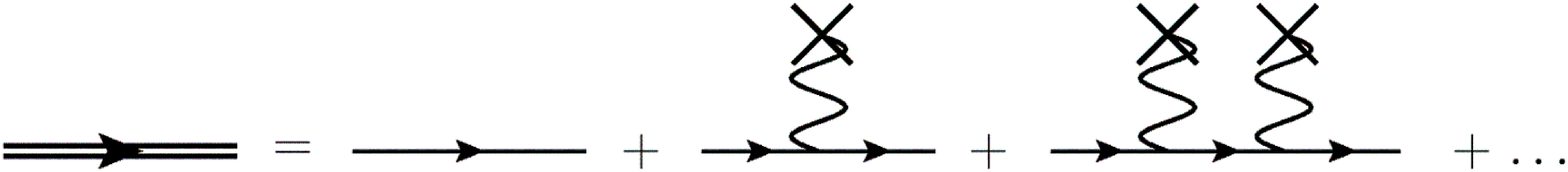}
\caption{The upper part depicts a diagrammatic representation of the beta function for the dimensionless coupling $g$ (red dot). 
 The regularized fermionic loop contains the full $B$ dependence as given by the fermionic
 propagator coupling to the external gauge field $\mathcal{A}$ to all orders as shown in the lower part. The coupling to $\mathcal{A}$ is indicated by a 
 cross. The $\tilde{\partial}_{t}$ derivative acts on the regulator-dependent part of the propagators and generates diagrams with regulator insertions (for details, 
 see Appendix~\ref{App:rg}).  
 \label{fig:diagrams}
}
\end{figure}
In this model, the non-trivial fixed point also acts as a separatrix for different regimes of 
the flow. These two different regimes translate into different phases realized in the respective ground
states of the quantum field theory~\cite{Wetterich:1992yh,Rosa:2000ju}. For values $g<g_{\ast}$, the flow is toward a noninteracting theory,
while for $g>g_{\ast}$ the coupling diverges at some finite IR scale $k_{\mathrm{c}}$, 
indicating that the discrete chiral symmetry is spontaneously broken. Thus, $g_{\ast}$ can be interpreted as the 
critical value of $g$ corresponding to a chiral quantum phase transition. Whereas $g_{\ast}$ is generally non-universal, the universal (scheme-independent) critical exponents of this quantum phase transition can be determined quantitatively with very good agreement among the various methods \cite{Hands:1992be,Karkkainen:1993ef,Hofling:2002hj,Braun:2010tt}.  
The order parameter for chiral symmetry breaking is the so-called chiral condensate, i.e., the expectation value of the operator $\sum_{i=1}^{N_{\mathrm{f}}}\altbarspinor\altspinor$.
It can be convenient to introduce an order-parameter field $\sigma$ into the functional integral
by means of a Hubbard-Stratonovich transformation. However, for giving a simple renormalization group
picture of magnetic catalysis and also for avoiding technicalities, we stay within a pointlike, purely fermionic description. For finite flavor number, we expect our arguments to be reliable until the onset of chiral symmetry breaking, where
a proper description of the built up of a finite expectation value of the order parameter and the associated
collective fluctuations need to be taken into account.

To capture the effects of a magnetic field on the RG flow, we make the following ansatz for the
effective average action:
\be
\label{eq:fermionic_action} 
\Gamma_{k}[\barspinor,\spinor,\mathcal{A}]&& =  \\
& &\!\!\!\!\int_{x}\Biggl\{\sum_{j=1}^{\Nf}\barspinor \mathrm{i} \fslash{D}[\mathcal{A}]\,\spinor + \sum_{i,j=1}^{\Nf}\altbarspinor\altspinor\frac{\bar{g}}{2\Nf}\barspinor\spinor\Biggr\},\nn
\ee
where the coupling $\bar{g}$ is now to be understood as a function of the RG scale $k$.
Inserting this ansatz into the flow equation, we can project onto the flow of  the dimensionless coupling $g$ by taking
appropriate functional derivatives with respect to the spinor fields. The presence of the magnetic
field requires a slightly different approach in the derivation of the beta function, which is usually
performed for the corresponding momentum-space quantities and fields. One problem is that the vector potential
$\mathcal{A}(x)$ entering the quadratic part of the action renders the fermionic propagator a
translationally non-invariant function of the space-time coordinates. Taking gauge holonomy factors appropriately into account, it still becomes possible to study the ansatz in \Eqref{eq:fermionic_action}, which
essentially constitutes the zeroth order in a derivative expansion. This implies that the derivative structure
contained in the effective average action is the same as in the classical action. In Appendix~\ref{App:rg} we demonstrate that
\Eqref{eq:fermionic_action} leads to a consistent flow equation within such a derivative expansion. There we also show that the scale
dependence of a wave function renormalization entering the action in the form $Z_{\psi}\sum_{j=1}^{N_{\mathrm{f}}}\barspinor\mathrm{i}\fslash{D}[\mathcal{A}]\,\spinor$ 
in the present truncation is trivial, i.e., $Z_{\psi}\equiv 1$ like in standard flows for pointlike fermionic theories with chiral symmetry~\cite{Gies:2010st,Braun:2011pp}.

The beta function $\beta_{g}\equiv\partial_{t}g$ is derived in Appendix~\ref{App:rg}. It incorporates the coupling of the fermions to
the external field to all orders in $qB$, see Fig.~\ref{fig:diagrams}, and reads
using a Callan-Symanzik regulator
\begin{equation}\label{eq:magneticbetafunction}
\partial_{t}g=(d-2)g-\frac{1}{16 \pi}\left(\frac{N_{\mathrm{f}}\,d_{\gamma}-2}{N_{\mathrm{f}}}\right)
g^2\left[\sqrt{\frac{2}{b}}\zeta\left(\frac{3}{2},\frac{1}{2 b}\right)-2 b\right].
\end{equation}
Here, we have defined  the dimensionless magnetic field parameter
\begin{equation}
b=\frac{q B}{k^{2}}.
\end{equation}
In the limit $b\to0$, the correct results for vanishing field are
recovered, cf. \Eqref{eq:fourfermionflow} and
Appendix~\ref{App:tf}. The magnetic field defines a new scale, the
so-called magnetic length $l=1/\sqrt{qB}$. The inverse magnetic length
thus describes the typical energy and momentum scales associated with
fermions moving in a magnetic field. For strong magnetic fields,
the fermionic fluctuations that experience Landau level quantization are primarily
driven by the lowest Landau level [$n=0$, $\tau=-1$ (cf. Appendix~\ref{App:sp})]. A Landau level can
roughly be attributed an extent of $l$ in all spatial directions orthogonal to the magnetic field. The present quantum field
theory in $d=3$ is thus dimensionally reduced via the external field to essentially  a quantum mechanical problem ($d\to d-2=1$).

In the regime $l^{-1}\ll k$, the high-momentum quantum fluctuations on the scale $k$ are not affected
by the long-wavelength physics associated with the low-momentum scale
$l^{-1}$. From the initial UV scale $k=\Lambda$ down to the magnetic
scale $k=l^{-1}$, we expect the beta function to be very similar to
its zero field counterpart. However, as we evolve the system toward
the IR by sending the RG scale $k$ to zero, the IR fluctuations in the
regime $l^{-1}\gg k$ are dominated by magnetic-field effects.

\begin{figure}
\includegraphics[scale=0.3]{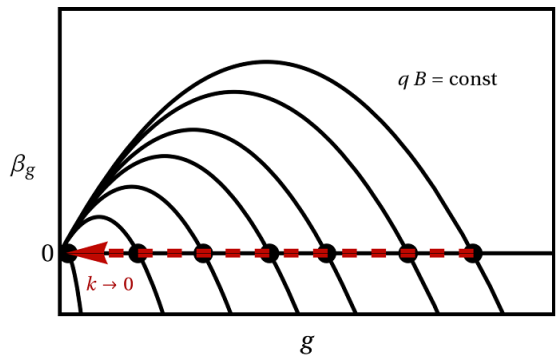}
\caption{Plot of the beta function $\beta_{g}$ (cf.~\Eqref{eq:magneticbetafunction}) of the coupling $g$ in the presence of
  an external magnetic field $B$ in arbitrary units.
  As the RG scale $k$ tends to zero for fixed $B$, the quadratic
  part of the beta function dominates over the linear part due to dimensional flow.
  The strong-coupling fixed point (black dots) is pushed toward the Gaussian fixed-point as
  the scale moves from the high UV to the IR regime (indicated by dashed red arrow). This leads to a divergence in $g$ at a
  finite scale associated with chiral symmetry breaking, for arbitrary values of the initial coupling
  at the UV scale $\Lambda$. From right to left $b$ takes the values $0.1$, $1$, $2$, $3$, $5$, $10$, $100$ and $1000$.
 \label{fig:parabolas}
}
\end{figure}

This behavior is depicted in Fig.~\ref{fig:parabolas}. At high scales ($b \gg 1$), the beta function shows 
the typical competition between a positive linear and negative quadratic contribution [cf. \Eqref{eq:fourfermionflow}]. Its
shape is unaffected by the presence of an external magnetic field. Upon lowering $k$, the non-Gaussian fixed point
starts moving toward the Gaussian fixed point. In the deep IR limit, the non-Gaussian fixed point asymptotically
merges with the Gaussian one. However, only the negative branch of the beta function survives in this limit, rendering
$g$ a relevant perturbation, completely oblivious of the initial UV value of the coupling. In this way, the divergence of $g$ 
at a finite scale is catalyzed by the external field for arbitrary values of $g$. 
 
A complementary approach to this phenomenon is the so-called ``quantum anomaly''~\cite{Gusynin:2005pk} for the Lagrangian
\be 
\mathcal{L}=\sum_{j=1}^{\Nf}\barspinor\mathrm{i}\fslash{D}[\mathcal{A}]\,\spinor.
\ee
It can be shown that the $\mathrm{U}(2 N_{\mathrm{f}})$ symmetry of this noninteracting system minimally coupled
to a perpendicular magnetic field is \emph{spontaneously} broken down to $\mathrm{U}(N_{\mathrm{f}})\times\mathrm{U}(N_{\mathrm{f}})$ for
$B\neq0$. One finds~\cite{Gusynin:1994va,Parwani:1994an,Dunne:1995cj,Dittrich:2000zu}
\be
\sum_{j=1}^{N_{\mathrm{f}}}\langle\barspinor\spinor\rangle=-N_{\mathrm{f}}\frac{q B}{2\pi}. 
\ee
However, for vanishing interaction ($\bar{g}=0$), this expectation value can not couple back into the bilinear part of the theory. So, although the chiral
symmetry is broken and a finite chiral condensate is dynamically generated, it does not act as a mass term for the fermions
and the spectrum of the theory remains unchanged.

\section{RG analysis of magnetically induced spectral gap}\label{sec:sg}

The phenomenon of spontaneous chiral symmetry breaking is accompanied by the
dynamical generation of a mass $\bar{m}_{\mathrm{f}}$ for the fermionic degrees of freedom which drive 
the symmetry breaking. In a partially bosonized formulation, which we already alluded to earlier,
the condensation of an order-parameter field $\sigma$ introduces a mass term into the free part of the
Dirac Lagrangian. The onset of symmetry breaking driven by quantum fluctuations typically occurs 
at some critical scale $k_{\mathrm{c}}$. We expect that $\bar{m}_{\mathrm{f}}\sim k_{\mathrm{c}}$.
The collective fluctuations of the fermionic system encoded in fluctuations of the $\sigma$ field about its expectation value will
typically diminish this value. In cases where a continuous symmetry is broken, the accompanying Nambu-Goldstone modes have the tendency to restore
the order, and the expectation value of the order-parameter field can be pushed to smaller values.

The beta function \Eqref{eq:magneticbetafunction} can now be used to analyze the field dependence of the dynamically induced fermion mass.
For this, we use the identification $\bar{m}_{\mathrm{f}}=k_{\mathrm{c}}$, ignoring possible $\mathcal{O}(1)$ factors arising from fluctuations in the symmetry-broken regime. This approximation is equivalent to a gap-equation approach \cite{Jaeckel:2002rm}. The critical scale $k_{\mathrm{c}}$ can be identified with the scale where the flow  of $g$ diverges. Therefore, it is convenient to reformulate the
beta function for $g$ as a beta function for the inverse coupling $1/g$. From $\beta_{1/g}$, we can move to the
actual trajectory $1/g(k)$ by solving the flow equation with a given initial condition $g(\Lambda)=g_{0}$ and record whenever $1/g(k_{\mathrm{c}})=0$. 

\begin{figure}
\includegraphics[scale=0.45]{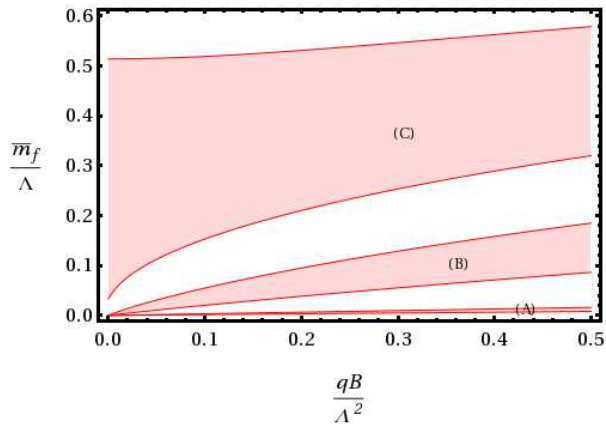}
\caption{$B$ dependence of the dynamically generated fermion mass $\bar{m}_{\mathrm{f}}$ as determined from
the critical scale $k_{\text{c}}$. The lowest boundary of each band corresponds to $N_{\mathrm{f}}=1$ and 
the upper boundary to $N_{\mathrm{f}}=100$. The curves for intermediate  $N_{\mathrm{f}}$ can be computed as easily, but are not shown for clarity. Band $(A)$ is the weak coupling $g_0\ll g_\ast$, band $(B)$ the intermediate but sub-critical coupling $g_0\lesssim g_\ast$, and $(C)$ the strong coupling regime, $g_0\gtrsim g_\ast$.
 \label{fig:regimes}
}
\end{figure}

The result of this analysis is shown in Fig.~\ref{fig:regimes}. The initial or bare value $g_{0}$ was chosen from three different regimes. The flow equation
was evaluated for flavor numbers ranging from $N_{\mathrm{f}}=1$ to $N_{\mathrm{f}}=100$, which at this level of the truncation
is indiscernible from the large-$N_{\mathrm{f}}$ limit. For clarity, only the curves for $N_{\mathrm{f}}=1$ and $N_{\mathrm{f}}=100$ are shown.
The curves for $1<N_{\mathrm{f}}<100$ fall into the shaded regions and do not cross in the $B$-interval we scanned. The precise values for $g_{0}$ and $\Lambda$ are listed in
Table~\ref{tab4}. In the weak coupling regime $g_{0} \ll g_{\ast}$ we find $\bar{m}_{\mathrm{f}}\propto B$. For intermediate, but sub-critical values of the initial coupling
$g_{0}\lesssim g_{\ast}$ deviations from the linear behavior show up. In a sense, in this regime the generation of a fermion mass is due to the
interplay of the `quantum anomaly' of the free system for $B\neq0$ mentioned in Sect.~\ref{sec:beta} and the four-point vertex $g$ (in the pointlike approximation).
While the anomaly generates a finite chiral condensate, the four-point vertex allows for coupling this condensate back into
the single particle spectrum, such that a mass gap opens up. In the strong coupling regime $g\gtrsim g_{\ast}$ the fermion mass is generated already for
$B=0$ in the usual chiral quantum phase transition from a massless phase to a phase with massive Dirac fermions controlled by the beta function for vanishing field.

%

\begin{table}[h!]\center
\begin{tabular}{p{50pt}||p{40pt}|p{40pt}|p{40pt}|p{40pt}p{0pt}}
\centering Regime  & \centering $g_{0}^{(\mathrm{RG})}$ & \centering $\Lambda^{(\mathrm{RG})}$ & \centering $g_{0}^{(\mathrm{DS})}$ & \centering $\Lambda^{(\mathrm{DS})}$ & \\ \hline\hline
\centering (A) $g_{0} \ll g_{\ast}$ & \centering $0.20$ & \centering $100\text{ a.u.}$ & \centering $0.206$ & \centering $100\text{ a.u.}$ & \\ \hline
\centering (B) $g_{0}\lesssim g_{\ast}$ & \centering $2.00$ & \centering $100\text{ a.u.}$ & \centering $2.76$ & \centering $100\text{ a.u.}$ & \\ \hline
\centering (C) $g_{0}\gtrsim g_{\ast}$ & \centering $6.50$ & \centering $100\text{ a.u.}$ & \centering $65.00$ & \centering $100\text{ a.u.}$ & 
\end{tabular}
\caption{Initial or bare value of the interaction $g_{0}$ and cutoff scale $\Lambda$ in the large-$N_{\mathrm{f}}$ RG and Dyson-Schwinger gap equation. The
scheme dependent critical values are $g_{\ast}^{(\mathrm{RG})}=\pi$ and $g_{\ast}^{(\mathrm{DS})}\simeq 5.39$.}
\label{tab4}
\end{table}

\begin{figure}
\includegraphics[scale=0.45]{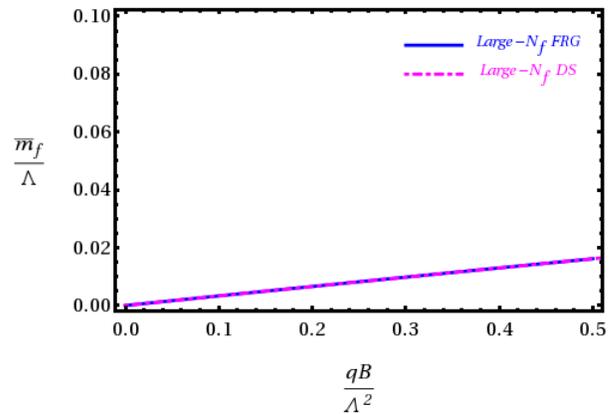}
\caption{Quantitative comparison of the $B$ dependence of the dynamically generated 
 fermion mass from the large-$N_{\mathrm{f}}$ flow equation (solid blue) and the large-$N_{\mathrm{f}}$ gap
 equation in the weak-coupling regime. For cutoff $\Lambda$ and bare value of $g_{0}$, see Table~\ref{tab4}.
 \label{fig:compA}
}
\end{figure}

\begin{figure}
\includegraphics[scale=0.45]{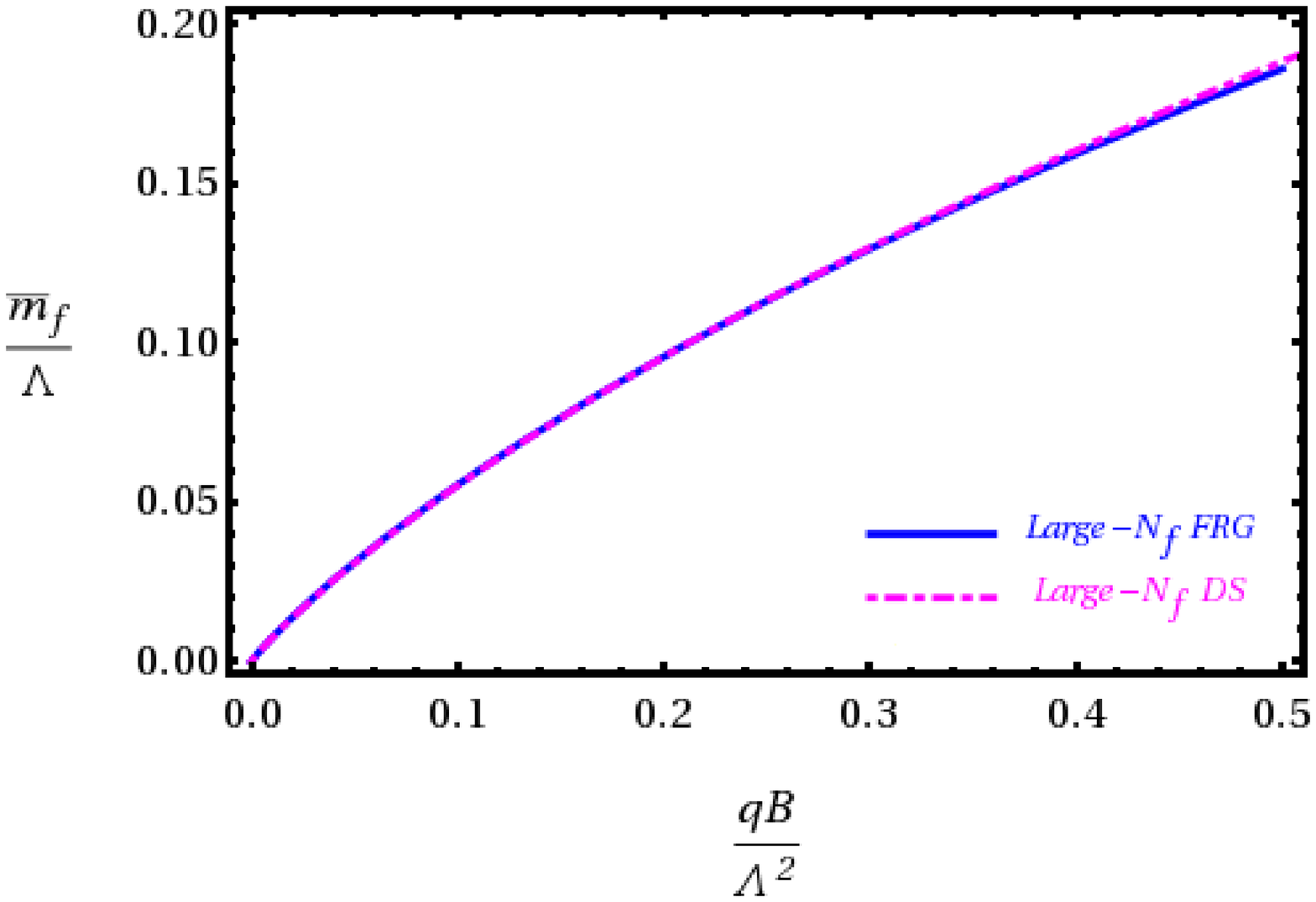}
\caption{Quantitative comparison of the $B$ dependence of the dynamically generated 
 fermion mass from the large-$N_{\mathrm{f}}$ flow equation (solid blue) and the large-$N_{\mathrm{f}}$ gap
 equation (dashed-dotted, magenta) in the intermediate to strong coupling regime. The coupling is still sub-critical. 
 For cutoff $\Lambda$ and bare value of $g_{0}$, see Table~\ref{tab4}.
 \label{fig:compB}
}
\end{figure}

\begin{figure}
\includegraphics[scale=0.45]{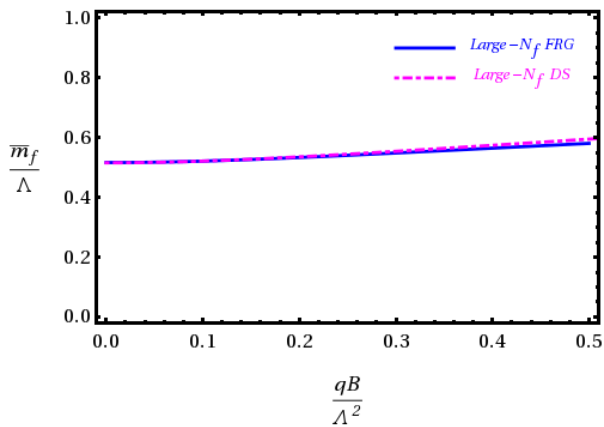}
\caption{Quantitative comparison of the $B$ dependence of the dynamically generated 
 fermion mass from the large-$N_{\mathrm{f}}$ flow equation (solid, blue) and the large-$N_{\mathrm{f}}$ gap
 equation (dashed-dotted, magenta) in the strong coupling regime. 
 For cutoff $\Lambda$ and bare value of $g_{0}$, see Table~\ref{tab4}.
 \label{fig:compC}
}
\end{figure}

To assess the validity of the pointlike approximation of the four point vertex and the naive identification of the dynamically generated fermion mass $\bar{m}_{\mathrm{f}}$
with the critical scale $k_{\mathrm{c}}$ we use in this work, we compared the large-$N_{\mathrm{f}}$ limit of our flow equation with the
large-$N_{\mathrm{f}}$ gap equation. This equation was derived in e.g.~\cite{Gusynin:1994va} and can be written as
\be\label{eq:gapequation} 
1=\frac{\bar{g}}{2 \pi^{\frac{3}{2}}}\int_{1/\Lambda^{2}}^{\infty}\frac{\mathrm{d}t}{t^{\frac{3}{2}}}\mathrm{e}^{-\bar{m}_{\mathrm{f}}^{2}t}(q B t)\coth\left(q B t\right).
\ee
The integral is regularized in the proper-time formulation and can be performed by analytic continuation, such that a closed
but cutoff-dependent expression is obtained (see~\cite{Gusynin:1994va}). The results are shown in Figs.~\ref{fig:compA}-\ref{fig:compC}. The values of the cutoff scale $\Lambda$ and the bare
interaction $g_{0}$ used in the two different schemes (RG and Dyson-Schwinger gap equation) are given in Table~\ref{tab4}. In order to compare these 
two approaches, the values of $\Lambda$ and $g_{0}$ were chosen, such that
the dynamically generated fermion mass $\bar{m}_{\mathrm{f}}$ obtained from the RG flow and the self-consistent solution to \Eqref{eq:gapequation}
coincide for $q B/\Lambda^{2}=0.1$. Since the Gross-Neveu model (or its corresponding universality class) only has a single parameter (which is true beyond
the pointlike approximation), this fixing prescription is sufficient. 

We find excellent agreement over a wide range of $q B$. Deviations occur due to cutoff effects when $q B/\Lambda^{2}\sim0.35$, as expected for scheme dependencies. Unfortunately, this quantitative check
cannot be carried over to finite $N_{\mathrm{f}}$ in a fully controlled manner. Especially the identification $\bar{m}_{\mathrm{f}}=k_{\mathrm{c}}$ cannot be expected to hold any longer,
since collective excitations -- which are completely damped in the large-$N_{\mathrm{f}}$ limit -- will modify the value of $\bar{m}_{\mathrm{f}}$ beyond
the critical scale $k_{\mathrm{c}}$ that for finite flavor numbers signals only the onset of chiral symmetry breaking. We also expect the strength of the transition to be weakened upon inclusion of order-parameter fluctuations, as has been recently observed in 4d quark-meson models \cite{Andersen:2011ip}.

%

\section{Conclusions and outlook}\label{sec:cao}

In this paper, we have studied the phenomenon of magnetic catalysis within the
three-dimensional Gross-Neveu model. Minimally coupling the Dirac fermions
to an external magnetic field catalyzes chiral symmetry breaking for arbitrary
bare values of the fermionic interaction. Even in the free system, chiral symmetry breaking occurs and
can be attributed to a form of `quantum anomaly'. However, only in the interacting case
does the symmetry breaking lead to a mass generation for the fermionic degrees of freedom. Using the
functional RG flow equation for the generating functional of 1PI vertices, 
we have obtained the beta function for the coupling parametrizing the fermionic interaction in a pointlike approximation. 
An analysis of the IR behavior of the beta function yielded a clear picture of magnetic catalysis
in the language of the renormalization group. The strong-coupling fixed point is pushed toward the Gaussian fixed point. But
in the deep IR limit, we find that only the negative branch of the beta function remains, and subsequently 
the coupling always flows into the strong coupling regime and eventually signals the breakdown of chiral symmetry
in a divergence at a finite critical scale.

This fixed-point picture of the approach to criticality in fact occurs in many fermionic systems \cite{Braun:2011pp}. For instance, also a gauge-field background giving rise to a non-vanishing Polyakov loop pushes the strong-coupling fixed point toward the Gaussian fixed point \cite{Braun:2011fw}. The unique feature of a magnetic background is the occurrence and subsequent IR dominance of the lowest Landau level acting as a symmetry-breaking catalyzer \cite{Gusynin:1994va,Gusynin:1994xp}.

By a symmetry analysis we provide insight into the structure of the Gross-Neveu theory space and more
importantly how additional magnetically induced operators might affect chiral symmetry breaking in external fields.
We find that a fermionic tensor-scalar contribution is generated for finite flavor number, when 
starting from the naive Gross-Neveu action with only scalar-scalar interactions.
It is subject to current investigation to study quantitatively the impact of these induced operators
on the value of the chiral condensate and the dynamically generated fermion mass in chirally symmetric quantum field theories.
The occurrence of these magnetically-induced operators is not particular to the Gross-Neveu model but certainly a generic feature of any fermionic system. In fact, this is yet another example for the proliferation of operators in the presence of further Lorentz tensors as is familiar from the classification of operators, e.g.,  at finite temperature \cite{Braun:2009si,Braun:2011pp}. It should be emphasized that mean-field studies typically ignore  contribution of such terms.

 Within our truncation,
there is no critical magnetic field, i.e., symmetry breaking occurs for all $B$. In principle, however, once further interaction channels
have opened up, there might be the possibility to compensate the `driving force' from the scalar-scalar channel, such that
for certain regimes, a critical field $B_{\mathrm{c}}$ might be necessary to enter the broken phase. This would also imply that the
interacting theory is not connected to the free theory in an analytic way as our present analysis suggests, and new scaling laws could
be observed.

Also, we neglected the back-reaction of collective neutral excitations onto the fermionic system. For small flavor numbers, we expect sizable quantitative modifications from the fluctuations of the order-parameter field. For infinite flavor numbers, we recover the
exact results of the large-$N_{\mathrm{f}}$ theory. The derivation of a consistent set of renormalization group
equations for the case of charged fermionic and neutral collective excitations is a work in progress and will be presented elsewhere.

Concerning condensed matter applications, precise quantitative control over the field dependence of the fermionic single-particle gap
might provide understanding of the interplay between magnetic catalysis and interaction-driven anomalous Hall plateaus in quantum Hall
measurements.

\acknowledgments The authors thank J. Braun, L. Janssen and
  J.M. Pawlowski for valuable discussions and collaboration on related
topics and acknowledge support by the DFG under grants Gi~328/5-1
(Heisenberg program), GRK1523 and FOR 723.


\appendix

\section{Spectrum of Dirac Operator in an External Magnetic Field}\label{App:sp}

Here we briefly summarize essentials of the spectrum of the square of the operator $\mathrm{i}\slash{D}[\mathcal{A}]$,
such that the trace appearing in the fermion loop (cf. Fig.~\ref{fig:diagrams}) can be conveniently expressed in the corresponding
basis. With the conventions in Sec.~\ref{sec:gn}, the squared Dirac operator can be written as
\be
(\mathrm{i}\slash{D}[\mathcal{A}])^{2} & = & -\left(D[\mathcal{A}]^{2}\mathds{1}_{4}-\frac{q}{2}\sigma_{\mu\nu}\mathcal{F}_{\mu\nu}\right)\nn\\
 & = & -\left(D[\mathcal{A}]^{2}\,\tau_{0}\otimes\tau_{0}+ q B\,\tau_{0}\otimes\tau_{3}\right),
\ee
where $D[\mathcal{A}]^{2}$ is the covariant Laplacian.
The spectrum of the squared Dirac operator is thus found to be
\be\label{eq:spectrum} 
\left\{p_{0}^2+\epsilon_{n}^{2}+\tau q B\,|\,p_{0}\in\mathds{R},n\in\mathds{N},\tau=\pm1\right\}.
\ee
Here, $\epsilon_{n}^{2}=|qB|(2n+1)$ corresponds to Landau level energies and $n$ is the associated Landau level
index. Accounting for the density of states $\frac{q B}{2\pi}$ for each Landau level,
the trace operation can be decomposed as
\be\label{eq:measure}
\mathrm{Tr}(\,\cdot\,)=\Omega \sum_{n\in\mathds{N}}\left(\frac{q B}{2\pi}\right)\int\nolimits_{-\infty}^{\infty}\frac{\mathrm{d}p_{0}}{2\pi}\,\mathrm{tr}(\,\cdot\,),
\ee
where $\Omega$ denotes space-time volume and $\mathrm{tr}$ is the trace over spinor indices.

\section{Derivation of the beta Function}\label{App:rg}

It is convenient to derive the flow equations in position space rather than momentum space due to the
presence of the vector potential $\mathcal{A}(x)$. Strictly speaking, the propagator is in general no longer a translationally invariant function of
space-time coordinates. It can, however, be decomposed into a holonomy factor times a translationally invariant factor~\cite{Dittrich:2000zu}, allowing in principle for a momentum-space formulation. Still, a formulation
in  position space is more direct in our purely pointlike approximation. The flow equation \Eqref{flowequation} can be expanded as
\be\label{eq:flowexpand}
\partial_t\Gamma_k & = & \frac{1}{2}\mathrm{STr}\left\{\left[\Gamma^{(2)}_k+R_k\right]^{-1}(\partial_tR_k)\right\}\\
& = & -\frac{1}{2}\mathrm{Tr} \tilde{\partial}_{t}\ln G_{k}^{-1} - \frac{1}{2}\mathrm{Tr} \tilde{\partial}_{t}\sum_{l=1}^{\infty}\frac{(-1)^{l+1}}{l}(G_{k} \tilde{\Gamma}_{k}^{(2)})^{l}\nn,
\ee
where the $\tilde{\partial}_{t}$ derivative only acts on the regulator dependent part.
Here we have decomposed the fluctuation kernel $\Gamma_{k}^{(2)}=\bar{\Gamma}_{k}^{(2)}+\tilde{\Gamma}_{k}^{(2)}$, where $\bar{\Gamma}_{k}^{(2)}$ is the
field-independent part and $\tilde{\Gamma}_{k}^{(2)}$ denotes the field-dependent vertex part. We also defined the scale-dependent regularized propagator
\be
G_{k}\equiv\left[\Gamma^{(2)}_k+R_k\right]^{-1}.
\ee
We choose a chirally invariant regulator 
\be 
R_{k}=Z_{\psi}\left[\mathrm{i}\fslash{D}[\mathcal{A}]r\left(\frac{(\mathrm{i}\fslash{D}[\mathcal{A}])^{2}}{k^{2}}\right)\right],
\ee
with an as yet unspecified shape function $r(x)$.
The fluctuation kernel can conveniently be derived in Nambu representation for the spinor fields (see e.g.~\cite{Gies:2001nw}). We thus obtain
\be
G_{k}=
\begin{pmatrix}
0 & G_{k}^{+} \\
G_{k}^{-} & 0
\end{pmatrix}
\ee
with $G_{k}^{+}=[Z_{\psi}(\mathrm{i}\slash{D}[\mathcal{A}])+R_{k}]^{-1}$, $G_{k}^{-}=[G_{k}^{+}]^{T}$ and
\be
\tilde{\Gamma}_{k,ij}^{(2)}=
\begin{pmatrix}
\bar{H}_{ij} & -F_{ij}^{T} \\
F_{ij} & H_{ij} 
\end{pmatrix}.
\ee
While we have omitted flavor indices for the propagator part, which is diagonal in flavor space, the flavor structure of the vertex part
is non-trivial. The matrix blocks are given by
\be 
\bar{H}_{ij} & = & -\frac{\bar{g}}{N_{\mathrm{f}}}\,\altbarspinor^{T}\barspinor,\\
-F_{ij}^{T} & = & \frac{\bar{g}}{N_{\mathrm{f}}}\left\{\altbarspinor^{T}\spinor^{T}-\sum\nolimits_{l=1}^{N_{\mathrm{f}}}(\bar{\psi}_{l}\psi_{l})\delta_{ij}\mathds{1}_{4}\right\},\\
F_{ij}^{T} & = & \frac{\bar{g}}{N_{\mathrm{f}}}\left\{\altspinor\barspinor+\sum\nolimits_{l=1}^{{N_\mathrm{f}}}(\bar{\psi}_{l}\psi_{l})\delta_{ij}\mathds{1}_{4}\right\},\\
H_{ij} & = & -\frac{\bar{g}}{N_{\mathrm{f}}}\,\altspinor\spinor^{T}.
\ee
We define the position space representation of the operator $\mathrm{i}\slash{D}[\mathcal{A}]$ as
\begin{equation} 
\langle x|\mathrm{i}\fslash{D}[\mathcal{A}]|x^{\prime}\rangle\equiv
\mathrm{i}\fslash{D}[\mathcal{A}](x,x^{\prime})=\gamma_{\mu}(\mathrm{i}\partial_{\mu}^{x}+\mathcal{A}_{\mu}(x))
\delta(x-x^{\prime}).
\end{equation}
Accordingly, we proceed for the regulator part $R_{k}$ inserted into the Gaussian measure of the generating functional of Green's functions:
\begin{equation} 
\langle x|R_{k}|x^{\prime}\rangle\equiv R_{k}(x,x^{\prime})=Z_{\psi}\left[\mathrm{i}\fslash{D}[\mathcal{A}]r\!\left(\frac{(\mathrm{i}\fslash{D}[\mathcal{A}])^{2}}{k^{2}}\right)\!\right]\!(x,x^{\prime}).
\end{equation}
In this notation, the regularized propagator is given by
\begin{equation}
G_{k}^{+}(x,x^{\prime})=Z_{\psi}^{-1}\left[\mathrm{i}\fslash{D}[\mathcal{A}]\left(1+r\!\left(\frac{(\mathrm{i}\fslash{D}[\mathcal{A}])^{2}}{k^{2}}\right)\right)\right]^{-1}\!\!(x,x^{\prime}).
\end{equation}
However, the wave function renormalization is $Z_{\psi}=1$ in this pointlike truncation. This can be
seen from the tadpole diagram giving rise to self-energy corrections. The regularized tadpole diagram is proportional to
\be 
\int_{x}\tilde{\partial}_{t}\,\mathrm{tr}G^{+}(x,x).
\ee 
Since it can be shown that the above mentioned holonomy factor becomes trivial for $x^{\prime}=x$,
the remaining contribution to the trace comes from the translation invariant part and thus the
tadpole carries no net momentum at all, like in the $B=0$ case.

The vertex part of the fluctuation matrix becomes in position space representation
\begin{widetext}
\be
\bar{H}_{ij}(x,x^{\prime}) & = & -\frac{\bar{g}}{N_{\mathrm{f}}}\,\altbarspinor^{T}(x)\barspinor(x)\delta(x+x^{\prime}),\\
-F_{ij}^{T}(x,x^{\prime}) & = & \frac{\bar{g}}{N_{\mathrm{f}}}\left\{\altbarspinor^{T}(x)\spinor^{T}(x) - \sum\nolimits_{l=1}^{N_{\mathrm{f}}}(\bar{\psi}_{l}\psi_{l})(x)\delta_{ij}\mathds{1}_{4}\right\}\delta(x-x^{\prime}),\\
F_{ij}^{T}(x,x^{\prime}) & = & \frac{\bar{g}}{N_{\mathrm{f}}}\left\{\altspinor(x)\barspinor(x)+\sum\nolimits_{l=1}^{{N_\mathrm{f}}}(\bar{\psi}_{l}\psi_{l})(x)\delta_{ij}\mathds{1}_{4}\right\}\delta(x-x^{\prime}),\\
H_{ij}(x,x^{\prime}) & = & -\frac{\bar{g}}{N_{\mathrm{f}}}\,\altspinor(x)\spinor^{T}(x)\delta(x+x^{\prime}).
\ee
\end{widetext}
The projection onto the beta function of the dimensionless coupling $g=k^{d-2}\bar{g}$ is facilitated
by collecting all contributions on the right-hand side of \Eqref{eq:flowexpand}, which are quartic in the fermionic fields. We find two contributions,
$\frac{1}{2}\mathrm{Tr} \tilde{\partial}_{t}\left\{G_{k}^{+} F G_{k}^{+} F\right\}$ and $\frac{1}{2}\mathrm{Tr} \tilde{\partial}_{t}\left\{G_{k}^{+} H G_{k}^{-} \bar{H}\right\}$.
These subsequently will be evaluated for spatially constant spinor fields. Using the identity
\be 
\tilde{\partial}_{t}G_{k}^{+}=-G_{k}^{+}\partial_{t}R_{k}G_{k}^{+},
\ee
and after some algebra, we arrive at
\begin{widetext}
\be\label{eq:flow_derivation}
\frac{1}{2}\mathrm{Tr} \tilde{\partial}_{t}\left\{G_{k}^{+} F G_{k}^{+} F + G_{k}^{+} H G_{k}^{-} \bar{H}\right\} & = &
\frac{2}{N_{\mathrm{f}}}\bar{g}^{2}\,\sum_{i,j=1}^{N_{\mathrm{f}}}\int_{x}\altbarspinor\left(\left[\mathrm{i}\fslash{D}[\mathcal{A}]\left(1+r\left(\frac{(\mathrm{i}\fslash{D}[\mathcal{A}])^{2}}{k^{2}}\right)\right)\right]^{-3}\partial_{t}R_{k}\right)(x,x)\,\altspinor
(\barspinor\spinor)\nn \\
& & -\bar{g}^{2}\mathrm{tr}\int_{x}\left(\left[\mathrm{i}\fslash{D}[\mathcal{A}]\left(1+r\left(\frac{(\mathrm{i}\fslash{D}[\mathcal{A}])^{2}}{k^{2}}\right)\right)\right]^{-3}\partial_{t}R_{k}\right)(x,x)
\sum_{i,j=1}^{N_{\mathrm{f}}}(\altbarspinor\altspinor)(\barspinor\spinor),\nn\\
\ee
\end{widetext}
where $\mathrm{tr}$ denotes the trace over spinor indices. The first term on the right-hand side in the first line of \Eqref{eq:flow_derivation} contains not only a contribution to the flow
of $g$, but also a term $\sim(\altbarspinor\sigma_{\mu\nu}\mathcal{F}_{\mu\nu}\altspinor)(\barspinor\spinor)$ is generated in an
infinitesimal RG step. Nevertheless, in the large-$N_{\mathrm{f}}$ limit, this contribution drops out from the fermionic loop.
It is also important to note that the spinors $\altbarspinor$ and $\altspinor$ are to be contracted with the term in round brackets.
To see this more clearly, it is suitable to employ a Laplace transformation,
\be
f(x)=\int_{0}^{\infty}\mathrm{d}s\tilde{f}(s)\,\mathrm{e}^{-xs},
\ee
with $x=(\mathrm{i}\fslash{D}[\mathcal{A}])^{2}/k^{2}$, in order to handle the appearing traces. Since the
spectrum of $(\mathrm{i}\fslash{D}[\mathcal{A}])^{2}$ is known (cf. \Eqref{eq:spectrum}), we perform the trace with help of \Eqref{eq:measure} and 
obtain
\begin{widetext}
\be\label{eq:flow_derivation2}
\mathrm{Tr}^{\prime}\mathrm{e}^{-(\mathrm{i}\fslash{D}[\mathcal{A}])^{2}/k^{2} s}=\Omega\left(\frac{q B}{2\pi}\right)\left[\frac{k}{4 \sqrt{\pi s}\sinh{\left(\frac{qBs}{k^{2}}\right)}}\right]
\left[\cosh{\left(\frac{qBs}{k^{2}}\right)}\tau_{0}\otimes\tau_{0}+\sinh{\left(\frac{qBs}{k^{2}}\right)}\tau_{0}\otimes\tau_{3}\right],
\ee
\end{widetext}
where the prime indicates that the trace over spinor indices is not included. The last term in \Eqref{eq:flow_derivation2} makes the 
generation of a magnetically induced operator explicit. However, neglecting this contribution, the flow equation for $g$ can be
derived in closed form by specifying the shape function $r(x)$ as
\be 
r(x)=\sqrt{\frac{x+1}{x}}-1,
\ee
i.e., by using the Callan-Symanzik regulator. The Laplace transform is known in this case and reads 
\be
\tilde{f}(s)=-\frac{1}{2}\mathrm{e}^{-s}s.
\ee

Upon performing the Laplace integral, we finally arrive at the rather simple result
%
\begin{equation} 
\partial_{t}g=(d-2)g-\frac{1}{16 \pi}\left(\frac{N_{\mathrm{f}}\,d_{\gamma}-2}{N_{\mathrm{f}}}\right)
g^2\left[\sqrt{\frac{2}{b}}\zeta\left(\frac{3}{2},\frac{1}{2 b}\right)-2 b\right],
\end{equation}
%
where $b\equiv\frac{q B}{k^{2}}$ denotes the dimensionless external field and $\zeta(x,y)$ is the 
Hurwitz zeta function.

\section{Threshold Functions}\label{App:tf}

The regulator dependence of the flow equation is encoded in dimensionless threshold
functions which arise from 1PI diagrams describing fermionic quantum fluctuations. In this work we used a Callan-Symanzik regulator.
In the case of vanishing magnetic field, the beta function for the coupling $g$ is given by
\be
\partial_t g = (d - 2)g - 4 \left(\frac{N_{\mathrm{f}}\,d_{\gamma}-2}{N_{\mathrm{f}}}\right) v_d\,l_1^{F}(0)\, g^2\,,
\ee
with the threshold function 
\be
l_1^{F}(\omega)=\tilde{\partial}_{t}\int_{0}^{\infty}\mathrm{d}y\,y^{\frac{d}{2}-1}\frac{1}{(1+r(y))^{2}+\omega},
\ee
see e.g.~\cite{Gies:2010st,Berges:2000ew} for a derivation. For the Callan-Symanzik regulator, the threshold function evaluates to
$l_1^{F}(0)=\frac{\pi}{2}$. We checked that as $b\to0$ the beta function \Eqref{eq:magneticbetafunction} has the correct limit and the
results for vanishing field are recovered. This is easily seen by noting that
\be
\lim_{b\to0}\left[\sqrt{\frac{2}{b}}\zeta\left(\frac{3}{2},\frac{1}{2 b}\right)-2 b\right]=4.
\ee

\end{document}